# Soap bubbles on the surface of a liquid: a playful universe in 3+1 dimensions, a real universe in 4+1 dimensions

H. Bartholin [(1)] and B. Barbara [(2)]

(1) Honorary Professor of the University (hbartholin@orange.fr)
(2) Emeritus Director of Research, Néel Institute, CNRS, 25 Av. des martyrs, 38042, Grenoble Cedex 01 (bernard.barbara@grenoble.cnrs.fr)

## Abstract

Very simple experiments based on an analogy between two theories —gravitation and capillarity— allowed us to construct a 2-dimensional-space toy-universe (U2D) composed of the surface of a 3D water tank with floating soap bubbles and bubble clusters. The experiments performed with or without a "black hole" suggest that our visible universe (U3D) is composed of the 3D "surface" of a 4D universe essentially filled with a heavy fluid that, interestingly, possesses all the properties presently attributed to "dark matter". The observations of this U2D and their implications for the structure and evolution of our U3D are discussed openly without preconceptions, not to say naively. This analogy led us to propose a scenario reproducing several observations made in our U3D. It would also allow us to schematically investigate the evolution of portions of our visible universe (stars, black holes, galaxies) at easily available timescales (seconds, to days) in contrast with the billions of years required for direct observations. More basically, it suggests that the 4D volume of our universe, which is inaccessible to us, constitutes a huge tank of heavy fluid sitting "below" our 3D surface universe and this has several consequences such as an original interpretation of the event horizon of black holes or a simple answer to the problem of missing dark matter.

## 1-Introduction

Starting from an analogy between capillarity and gravitation we experimentally created —approximately 15 years ago— a model for a two-dimensional universe (U2D) and compared it to our own three-dimensional universe (U3D). This model simply consists of the observation of different static or dynamical configurations of soap bubbles floating on the surface of a tank filled with a liquid (water) which, through the capillary forces analogous to the Newton's gravitational forces, gather in aggregates that represent the massive objects of our U3D, such as planets and stars. After some improvements we also created a U2D black hole. The water menisci that form around each floating object —from smallest bubbles to large aggregates of bubbles— can be considered as analogous to the Einstein space-time curvatures around masses. This gravitation/capillarity analogy allowing us to reproduce in 2D many of the cosmological observations made in our U3D led us to interpret several observed phenomena in the context of a 4D universe (5D if time is taken into account).

## 2- The gravitation/capillarity analogy

It is well known that the attraction force between two objects floating on a $2D$ interface (e.g. liquid-air) such as two bubbles of capillary masses $Q_{1,2}$ distant by $R$ is proportional to $Q_1Q_2/R$. Its similarity with the Newtonian gravitational force between two masses $M_{1,2}$ proportional to $M_1M_2/R^2$ in $3D$ suggests, for both, a more general expression in $1/R^{D-1}$ where $D$ is a macroscopic spatial dimension equal to 3 in our gravitational space and to 2 in the capillary case. This analogy, shown in detail in the past (not published) in which a bubble – or any other floating object– plays the role of a mass in our universe, is reinforced by the fact that in both cases the attraction is linked to the recovery of small space curvatures around the masses: the liquid menisci in the vicinity of the floating objects (weak $3D$ curvatures of the $2D$ surface of the liquid) and the Einstein space-time curvatures in the vicinity of masses in the case of gravitation (weak $4D$ "space-time" curvatures of the $3D$ space).

## 3- Evolution of a carpet bubble to an ensemble of isolated bubbles and the formation of stars

Our starting point consisted, quite simply, of filling a container with a liquid, water for example, and adding at the same time a small amount of soap giving soap bubbles a result familiar to everyone. With some imagination on can say that the obtained configurations of bubbles look like what one can see in the sky with its configurations of stars, galaxies etc…

After finding the ideal composition of soapy material and liquid, we obtained an ensemble of bubbles that, due to their capillary interactions, self-organize slowly in the form of aggregates with fractal-like size distributions (Fig. 1). These configurations are similar to representations of the beginning of the evolution of a portion of galaxy of our U3D which, due to gravitational interactions also self-organizes from a homogeneous state into fractal distributions of matter (from star dust to stars) [1] [2], the 3D surface of which constitutes our observable universe.

Another type of bubble distribution that is obtained corresponds to the case where a self-organized configuration of the type shown Fig. 1 gradually gathers to form larger and larger independent clusters with the shape of discs (Fig. 2). Such an evolution, also results from the capillary interactions between bubbles. This progressive transformation takes time and requires an initial smaller density of bubbles (i.e. of soap) i.e. a sparser initial fractal distribution of bubbles. In our analogy, this suggests that distant spherical stars and other celestial objects (analogues of the objects of Fig. 2) emerge in regions of the "universe" (here of Fig. 1) which are not extremely dense. Indeed, in our model, while extremely dense regions give rise to high concentrations of bubbles and of their clusters (e.g. large stars close to each other in our universe), more dilute regions give rise to the distant "stars" of medium sizes of Fig. 2.

In short these two figures simulate, in a very schematic way, the transformation of the initial diffuse matter of a portion of the universe into a self-organized region with a fractal distribution of mass (Fig. 1) or into a region with well defined objects such as stars and then in galaxies (large independent bubbles clusters with comparable sizes, here), rather distant from each other (Fig. 2). In this capillary/gravitational similarity, the size distributions of the obtained stars depend on the initial conditions, such as the initial density of soap/matter.

The above examples correspond to the case where the gravitational forces are dominant, meaning that we are in a region of space without a black hole. This picture should of course, be completely disrupted in the presence of a black hole. This is what will be considered now. Before looking at the amazing internal structure of black holes revealed by our analogy, we shall first consider how they affect their close environment.

### 4- Effects of a black hole on surrounding matter and galaxies

Even if black holes result from the gravitational collapse of stars in regions with extremely high density of matter, it is obviously no question to simulate a U2D black hole by increasing the concentration of bubbles in our water tank. Indeed gravitational collapse involves extremely intense reactions such as thermonuclear fusion and even more with, obviously, no capillary equivalent. In fact, we took another, very simple solution: dig a hole at the bottom of our water tank (Fig. 3).

We can see (Fig. 4) surface water motions around point Os situated at the vertical direction of the hole. These movements known to all, which transform themselves into whirls and tourbillons absorbing the bubbles and clusters of bubbles (representing stars, planets, etc.) constitute the black hole of our U2D. This "black hole" appearing at the center of Fig. 4, is surrounded at larger distances, by its 2D "spiral galaxy".

To show more clearly how the matter is absorbed by the "black hole" a more extended view of the black hole (in Os) with surrounding bubbles and clusters of bubbles is given Fig. 5. All these floating objects rotate and deform themselves according to spirals ending at point Os, being peeled bubble by bubble or small cluster by small cluster. Note that due to their strong internal capillary forces making them more compact, the largest clusters rotate with longer periods of time which, however, become faster as they move closer to point Os before being also dragged and absorbed. These pictures accurately represent the way real black holes absorb matter in our universe. In particular star peeling constitutes a well-known mechanism for the absorption of matter by black holes [3]. Interestingly, one can see that such a rotational movement of matter far around the central point Os may constitute an indirect mechanism for the detection of a central black hole that would be invisible due, for example, to a complete lack of matter in its close environment. Indeed, if a black hole absorbed all the matter around without being fed it would no longer be visible, the surface gravitational fluid that it carries, without matter, was not visible (dark blue point on the second half of the images of Fig. 5, for details, see section 5). After a sufficiently long time, when bubbles and clusters of bubbles (matter, stars in our U3D) start again to gravitate around the black hole and be absorbed by it the black hole should gradually become increasingly visible. After even more time a new galaxy of bubbles can be regenerated (returned to the first images of Fig. 5). These U2D observations provide a simple possible interpretation for the successive appearance and disappearance of black holes, and even of galaxies. Research is underway on this topic, of the galaxy Abell 2261 [4].

Fig. 5 shows that after it has started to be peeled by the rotating "black hole" or has been split into two parts, an aggregate can rotate more quickly around Os. This should, in principle, allow an escape from the attraction zone gaining kinetic energy as suggested by Penrose [5] — a process based on Hill's mechanism [6] to extract energy from the expulsion of a star gravitating around a rotating black hole [7]. In general, we were not able to observe this effect after our aggregate was divided into two parts because its rotation speed was not fast enough, except in one case when the clusters were in fast rotation near Os. Nevertheless, we observed (Fig. 7) how two clusters rotating far enough from the point Os can meet during their approach to this item to form a single larger cluster. This phenomenon occurs when the clusters are close enough to each other so that their capillary forces are greater than their driving forces towards Os. This is analogous to the grouping of a double star sufficiently far from a black hole, as seen in Fig. 6 and Ref. [8].

Fig. 8 left shows how, in our 2D experiments, a galaxy can be born and move before it disappears, in the absence of a black hole as observed in our universe (Ref 8). This was obtained by observing the collective motion of an ensemble of bubbles and clusters of bubbles achieved after mixing water and soap locally for a few seconds. The 2D "galaxy" does not appear systematically but, when this is the case, in addition to its own overall rotation, it moves slowly following the random walk of its center. Such a galaxy, not linked to a hole can also be drained by other types of water currents(different from those initiated at the beginning of the experiment) due to the bubbles and aggregates sitting near the edges of the tank —which represent the motion of stars and other celestial objects sitting in the environment of a real galaxy. These conflicting currents ultimately lead to the disappearance of the galaxy due to the loss of its collective angular moment. We also observed that the usual type of galaxy —built around a "black hole"— observed in our experiments if we drill a hole at the bottom of the tank (Fig. 8 right, for details see next section) can also move, disappear, and even in this case, reappear (Ref. 9). This figure also shows that our 2D "black hole" sitting at the center of its "galaxy" moves in a region no longer vertical to the hole. This situation appears when the distance $h$ between the liquid surface and the hole is larger than a characteristic distance ($h>h_i$, see next section) for which the liquid surface is not strongly connected to the hole, allowing an oblique aspiration whose direction is sensitive to the parasitic currents circulating around the galaxy. This results in a random but limited, motion of the 2D "black hole" ("wandering black holes" [9]) taking place in very turbulent regions of our universe such as near supernovas or in the proximity of black hole collisions. Then, after waiting even more time the random water currents decrease in intensity to eventually disappear allowing the "galaxy" and its collective motion around its "black hole" to refocus near the vertical direction of the hole allowing the surface 2D "black hole" to reappear (Fig. 8, right, last image). Such a "reappearance" is observed when the distance $h$ decreases below the critical value $h_i$, suggesting that the observed mobility of a black hole may depend on its distance to its escape point (see next section). By the end of this paper, we shall see that we may provide other mechanisms explaining the similar appearance/disappearance of well-connected black holes when $h<h_i$.

## 5- Simulation of a black hole and its extra dimension

In this section we describe the geometry and behavior of our black-hole model. Fig. 3 and Fig. 9a show that the water playing the role of the 4D fluid and the bubbles playing the role of the "real matter", are discharged in Os through the attraction by the hole in O creating on the surface small eddies (when $h$ is large) which progressively transform into large whirlpools (when $h$ is smaller). In our analogy, these eddies and whirlpools are analogs of what a black hole produces at the 3D surface of our 4D space universe. If the central discs of Fig. 9 (white because of the important density of accumulated bubbles, reflecting the light) represent the central parts of the black holes, the more or less gray halo on their contours represents their accretion discs (Fig. 9b and 9c).

The fact that the density of matter of the halos decreases with $h$ (they become darker and darker) is due to an increasing transfer of matter from the halos to the central discs in the whirls. This effect takes place simultaneously with a sinking of the discs and, to a lesser extent, of the halo below the surface of the liquid, which increases when $h$ decreases (this effect was amplified for more visibility in Fig. 9). Note that the transfer of bubbles from the halo to the central part is due to the angular momentum of the bubbles, which dictates their motion as they rotate around the black hole. As collisions occurs between bubbles and small clusters of the halo, their angular momentum is redistributed, and they settle into a disk with less collision orientation. A similar redistribution of angular momentum also takes place in real black holes and their accretion discs, explaining their flat shape (see below).

Let us now come to the inner structure of black holes, as suggested by our analogy. Fig. 9 shows that, around the bowls, the 2D liquid surface presents a 3D negative curvature created by the whirls and tourbillons at the vertical direction of the point O. Following our analogy, the 3D contour of real black holes of our universe, made of heavy fluid and matter, should show a similar 4D negative curvature while their centers plunge more or less into the 4$^{th}$ dimension. Such immersions near or even into the 4$^{th}$ dimensions (depending on the value of the length $h$ in our experiments). Such negative curvatures of the heavy fluid on the contour of the central object (the black hole) recall the positive curvatures on the contour of bubbles (capillarity menisci) and of their equivalent positive curvatures on the contour of masses (Einstein gravitation "vacuum curvatures") suggesting a possible common origin in a generalized dynamical Einstein's space-time theory… In any case, the central part of black holes, belonging to a region close to the 4D space (or in it) should not be accessible to us except if the "bowl" is accidentally empty of 4D fluid (see below). Fig. 9 shows that this effect is less pronounced, when the distance $h$ is larger suggesting that the central part of black holes associated with large $h$ may be more easily visible than the others. If $h$ is even larger the black holes become "weak" and can move as discussed above (see also Fig. 7). Clearly, in our experiments the 3D earth gravity, acting on the bubbles of the 2D surface of our water tank, plays the role of the 4D gravity of the heavy fluid filling the inaccessible part of our universe acting on the masses of its 3D surface (our visible universe). Interestingly, this picture solves the problem of missing dark matter.

The simplest way to interpret the meaning of the distance $h$ (Fig. 9) would be to consider the existence of another 4D universe filled with the same "gravitational fluid" but with a weaker density/pressure —and so weaker gravitation and universal constants— to allow its transfer from the first universe to the second one. Then, $h$ is the distance along the 4$^{th}$ dimension, between the black hole sitting on the 3D surface of the first 4D universe (absorbing fluid and matter) and a "white hole" sitting on the 3D surface of the second 4D universe (spitting fluid and matter). This obviously recalls the — sometimes — accepted idea of parallel universes or multiverses eventually connected by macroscopic wormholes through instantaneous macroscopic entanglement, suggesting that our 4D "gravitational fluid" is in fact, a 4D superfluid as this is sometimes suggested [10]. In contrast, other types of possible black holes splitting matter in our universe instead of absorbing it, should be linked with a universe of larger fluid pressure and therefore stronger gravitation and universal constants. Other, less obvious consequences might be assessed such as the fact that the space dimensions of parallel universes, even if they should be the same as those of our universe, may nevertheless be different — larger or, less probably smaller. At first sight, one may think that when the distance $h$ becomes very small (last images of Fig. 9c) the two universes are very close to each other's a situation, which certainly does not happen often… However, this simulates the case where the whirls and the currents leading to the evacuation of liquid and bubbles are so intense that all the environmental matter concentrates at its center and spins very quickly transforming the black hole into a single cluster of very small bubbles (Fig. 9c). This situation appears to be the result of the collapse of a large star, with a high speed of rotation at the end of its life before shutdown which, continuously, loses matter to another universe. Therefore Fig. 9c shows that after all the matter around the black hole has been absorbed, a large white circle remains in rotation without the apparent structure characteristic of our Kerr type "black holes" (Fig. 9a and 9b). Instead, this white circle resembles a star of our previous figures even though its size, depending ofn the size of the hole at the bottom of the tank, may be much larger than our "stars" (bubbles and clusters of bubbles). Therefore, such a remaining of a black hole appearing before its disappearance, might be taken for a large "star" originating from a black hole and disappearing with it after all its matter has been engulfed in the hole. Interestingly, these 2D observations could be compared to those made in our U3D universe according to which a massive star of 25 solar masses (N6946-BH1a) was observed becoming a black hole by disappearing suddenly, without going through the usual supernovae (failed supernova). It seems admitted that such a transformation of massive stars into black holes occurs with thirty percent of massive stars [14]. However, if one follows our analogy, we would not say that there is a transformation of a black hole into a big star, which, finally disappears, but that we have from the beginning to the end a black hole which transforms itself at the end of its life in something which looks like a big star, and then disappears. However, our 2D experiments show that the bubbles constituting this "big star" lie at the bottom of an empty bowl. This suggests that such a "big star" would be surrounded by important space curvatures, as is the case for a black hole. This star may strangely resemble a supernova before shutdown [11]. However, in our context this "star" represents the remains of a black hole that ends up being absorbed by another universe. Therefore, the question arises: may a supernova emerge from a black hole disappearing into another universe?

In fact, our simple simulation of a black hole is similar to the famous rotating Kerr black holes with their "accretion disc" which represent most of the black holes observed in our universe. Note that, the 2D halo (Fig. 9) implies a 3D accretion disk for a real black hole (i.e. filling the space between two concentric spheres). However, the pictures of Kerr accretion disks instead show, toroidal ring-like structures, a difference that is easy to understand and attributed to the abovementioned redistribution of angular momentum due to collisions that transforms spherical halos into nearly planar halos.

Returning to Fig.9, the spinning velocity of the halos increases when $h$ decreases *i.e.* when an increasing number of bubbles are transferred to the central disc, they also result from the conservation of the total angular momentum (disc plus halo) as is the case for real black holes with their accretion discs [12]. This transfer of matter from the halo to the deeper central aggregate is accompanied by a decrease in the radius of the latter to the benefit of the halo. More precisely, as $h$ decreases the bowl (Fig. 9) increasingly deforms and sinks resulting in increased "verticalization" of its walls with subsequent decreased bubble density of the halo. When the walls of the bowl, almost vertical, are completely devoid of bubbles the halo turns black because it loses all its bubbles that are evacuated into the disc and the hole. The swallowing of the last small bubbles —not visible— scattered on the surface around Os, makes it possible to visualize the last tourbillon which will then be limited to the real dimensions of the hole at the bottom of the tank where the last bubbles are concentrated before being completely evacuated when $h \rightarrow 0$.

The analogy with our universe predicts the existence of very different types of black holes corresponding to the different steps shown in Fig. 9 i.e. on the 4D distance $h$ to their unloading point (equivalents to the distance $h$ to the hole at the bottom of the tank). In particular, the density of matter, the size of "black holes" hearths and their halos change drastically when $h$ decreases giving different appearances. These figures, as Fig. 4 also, provide a probable reason for a black hole to disappear after it has absorbed all the matter of its near environment while continuing to absorb the invisible 4D fluid. Clearly, this does not mean that it disappeared but just that it is not visible and would reappear if enough matter comes, from the surrounding galaxy, sufficiently close to it to be attracted and absorbed. The corresponding timescale, whatever it would be, should be much longer than in the case of the turbulences in the « bowl » mentioned above.

More generally, the different stages of black hole configurations observed in Fig. 9 seem to correspond, in our U3D, to different types of Kerr black holes [13]. In particular, the fact that the speed of rotation of the central disc representing the center of the black hole in Fig. 9a,b accelerates when $h$ decreases suggests that, in our universe, the speed of rotation of the central part of Kerr blacks holes could be different from each other's, depending on the distance to their assumed escape point. In addition, the diameter of this central aggregate being much larger than that of the real hole drilled at the bottom of the tank (which an idea of size is given by Fig. 9 when $h \rightarrow 0$, last schemes), suggests that what one sees of a black hole of our universe is much larger than its supposed escape point.

The last images of Fig. 9c show that after all the matter around the black hole has been absorbed, a large white circle remains in rotation without the apparent structure characteristic of our Kerr type "black holes" (Fig. 9a and 9b). Instead, this white circle resembles a star of our previous figures even though its size, depending on the size of the hole at the bottom of the tank, may be much larger than our "stars" (bubbles and clusters of bubbles). Such remaining of a black hole appearing before its disappearance, might be taken for a large "star" originating from a black hole and disappearing with it after all its matter has been engulfed in the hole. Interestingly, these 2D observations could be compared to those made in our U3D universe according to which a massive star of 25 solar masses (N6946-BH1a) was observed becoming a black hole by disappearing suddenly, without going through the usual supernovae (failed supernova). It seems admitted that such a transformation of massive stars into black holes occurs with thirty percent of massive stars [14]. However, if one follows our analogy, we would not say that there is a transformation of a black hole into a big star, which, finally disappears, but that we have from the beginning to the end a black hole which transforms itself at the end of its life in something which looks like a big star, and then disappears. However, there is a notable difference: in our 2D experiments the bubbles constituting this "large star" lie at the bottom of an empty bowl suggesting that such a "large star" would be surrounded by important negative space curvatures, as is the case for a black hole, with subsequent curved light trajectories.

Last remarks: (i) what we observe in our experiments, suggesting that the matter carried through a black hole escapes to another universe through a macroscopic wormhole — the role of which being, here, played by the earth — is none other than one of Hawkings' flagship ideas, (ii) if the $4^{th}$D distance between what we see of Kerr black holes on the 3D surface of our universe and their escape points should never be accessible to us, some of their characteristics such as the size of their central parts, the ones of their halo, their density of matter, their speed of rotation …. may give relative indications about the distances to their escape points, allowing a hierarchy for such distances and then, knowing the positions of each black hole on the 3D universe surface, some indications about the shape of our 4D universe.

The above analogy, between the effects of a hole at the bottom of a water tank and real black holes, seems to constitute a hydrodynamic extension of our initial capillarity/gravitation analogy. Indeed, in the two cases, curvatures of the heavy fluid take place on the object contour: (i) the weak positive 3D liquid curvatures of the menisci around the floating bubbles or clusters of bubbles (Fig. 9) analogous to the weak positive Einstein's 4D space-time curvatures around massive objects and (ii) the negative curvatures of the 2D surface of our 3D fluid around the central disc and halo of our "black hole" (centered at point Os, Fig. 4) analogous to what we suggest for real black holes with a strong negative 4D curvatures of the 3D surface of the heavy fluid in their contour (similar to the "bowls" of Fig. 9). The fact that the weak curvatures associated with capillarity/gravitation are both positive is consistent with the fact that these objects (e.g. bubbles/planets) attract each other. However the strong negative curvatures appearing in the inside face of the "bowl" of black holes result from the fact that they are sitting slightly below the 3D "surface" of the 4D fluid, sinking into the 4D space (Fig. 9). We could not determine whether such characteristics lead to an attraction or a repulsion of two black holes, but this could be done.

In addition, Fig. 9 also shows that the high rotational velocity of the fluid at the vertical direction of the hole transforms all the absorbed clusters of bubbles in a very dense carpet of the smallest possible bubbles sitting at the bottom of the "bowls", deeper and deeper in the 3D space when $h$ decreases. Therefore, according to our analogy a similar intense rotational gravitation of a real black hole should reduce the matter it absorbs to the state of an extraordinarily high density of rotating elementary particles plunging into the 4D heart of the black hole (our 3D "bowl") to disappear without any possibility of escaping from its 3D contour. This suggests that the "black hole horizon" i.e. the 3D surface at the center of a black hole from which nothing can escape sits at the limit between the 3D surface of our universe and its 4D black hole hearth where the strong 4D negative space curvatures becomes significant (3D in our model, Fig. 9). Therefore, our negative curvatures a simple continuation of the positives curvatures associated with the capillarity/Einstein-gravitation theories. It could be that these negative curvatures might be described by a dynamical generalization of Einstein's gravitation.

Finally it is clear that if the Einstein 4th dimension is directly associated with time and confined into our 3D space the 4th dimension of this paper is purely spatial even if it not excluded that time could intervene in an indirect way such as, as a result of an expansion of the whole 4D universe, which would be at the origin of the expansion of its 3D "surface", our universe.

## 6- And what about the creation of our universe?

The simplest image for our 4D universe would, obviously, be a 4D sphere similar to a 3D bubble of liquid which, in a weightless state, takes the form of a sphere "floating" in an infinite space of "pure vacuum" (i.e. completely empty, in opposition to quantum vacuum, space filled of particles gravitons, …forming our "heavy fluid"). A black hole is then similar to a crater sitting at a certain point of the 3D surface of this 4D sphere and plunging in the 4D volume that absorbs the fluid and the matter of its environmental 3D surface. As, a black hole can also emit matter instead of absorbing it, it is natural to consider it as a "white hole". If such a white hole belonging to another 4D spherical universe with a lower density of fluid (or, for us, more likely superfluid) were connected to a black hole of our universe, this should allow the fluid transfer into our universe, through the abovementioned macroscopic wormhole, here a 4D filament with a 3D section.

To extend this scenario to the creation of our universe (or of another one), one simply may assume that a wormhole issued from a 4D universe (and possibly entangled to it) ends at any point of the completely empty space (even without gravitational superfluid) to which the different 4D universes belong. The ejected jet of gravitational fluid should then take the shape of a rapidly growing or, say, exploding 4D sphere similar to what one could imagine if a black hole exploded and which is, roughly similar to a Big Bang. The extremely fast increase in the 4D radius of this "sphere" of heavy fluid should slow down with time because the pressures between the initial and the new universe tend to equilibrate. Therefore, the 3D surface of the created universe should also increase but less and less rapidly in time, as this is the case for our visible universe. This short discussion, relating to the creation of a universe, might connect with what we know of the evolution of the oldest observed black hole J1342+0928, a giant with nearly the age of our universe [14]. Clearly, this very schematic scenario suggesting how our universe might have started cannot enter into the already well-known details of the evolution of our universe after the Big Bang.

## 7- Conclusion

Based on an analogy between capillarity and gravitation, this paper proposes a scenario for the description and evolution of our universe using two types of very simple, but rigorously conducted experiments referring to (i) the behavior of celestial objects in our universe such as stars and galaxies and (ii) the behavior and structure of black holes. More precisely, the first type of experiment shows that the evolution of different types of configurations of ensembles of 2D bubbles and clusters of bubbles sitting at the 2D liquid/air interface of a 3D water tank mimics surprisingly well the evolution of celestial objects "floating" on the 3D surface of a 4D universe, our universe. The second type of experiment showing the effects created, on floating bubbles and clusters of bubbles, by a hole dig at the bottom of a water tank also simulates, surprisingly well, a black hole with its surrounding galaxy. This analogy suggests that, if black hole halos are always located on the 3D surface (our observable universe), their central parts can be, dynamically so dense that they may "pierce" this 3D surface to penetrate its 4D volume, and then become absolutely inaccessible to us. If this were the case, the event horizon of a black hole, would be simply defined as the limit between its 3D outside part and its 4D inside part with its enormous dynamical mass. Clearly enough, the water of our 3D tank absorbed by the hole represents a "gravitational fluid" filling our 4D universe which, interestingly, possess most of the properties attributed to the so-called "dark matter" (extremely heavy, accumulating around masses, inaccessible to us and to our instruments). However, it is clear that we could not determine the precise nature of this fluid except the fact that it is at the basis of gravitation and therefore that it must contain gravitons. An interesting extension of all that would be to perform the same type of experiments with a superfluid (such as $He_3$) instead of water to have access to a description of what remains at the macroscopic scale of a multiverse based on quantum gravitation and cosmology. Last but not least this paper written, as a game by non-specialists of the field, has no particular ambition except to show in a way voluntarily naive, how far can lead this analogy between capillarity and gravitation.

# Figures and Figure Caption

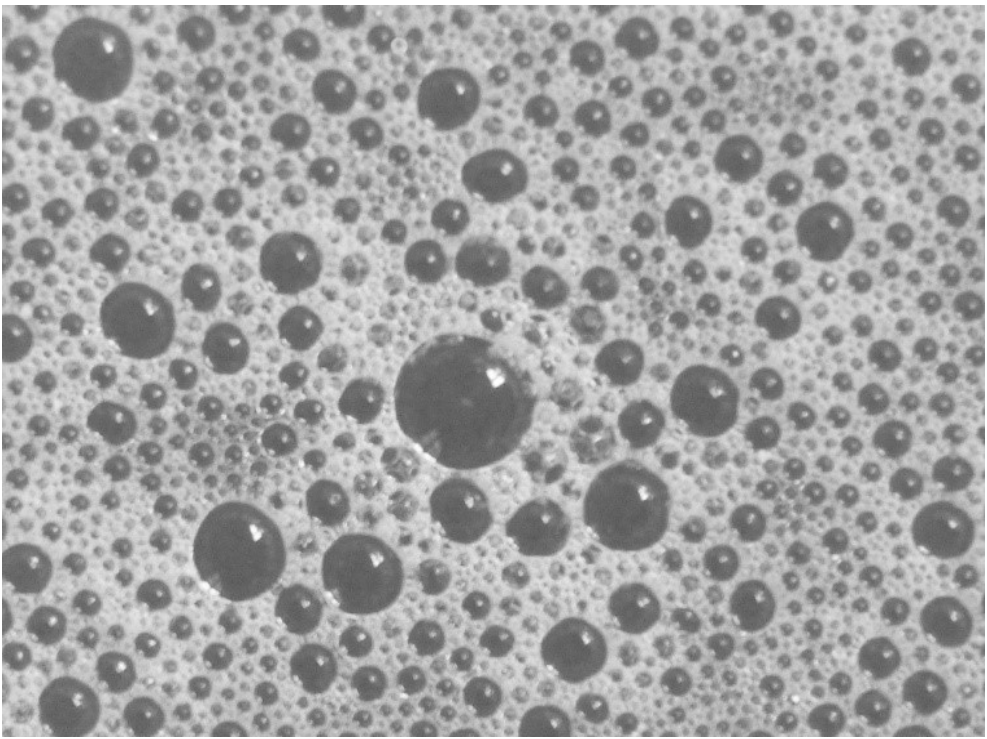

Fig. 1: Soap bubbles of all sizes, with fractal-like distributions, before they reach their equilibrium state. Due to their capillary interactions, the bubbles of a given size are surrounded by smaller and smaller bubbles. Clearly, the obtained mean density of bubbles decreases with the initial proportion of soap.

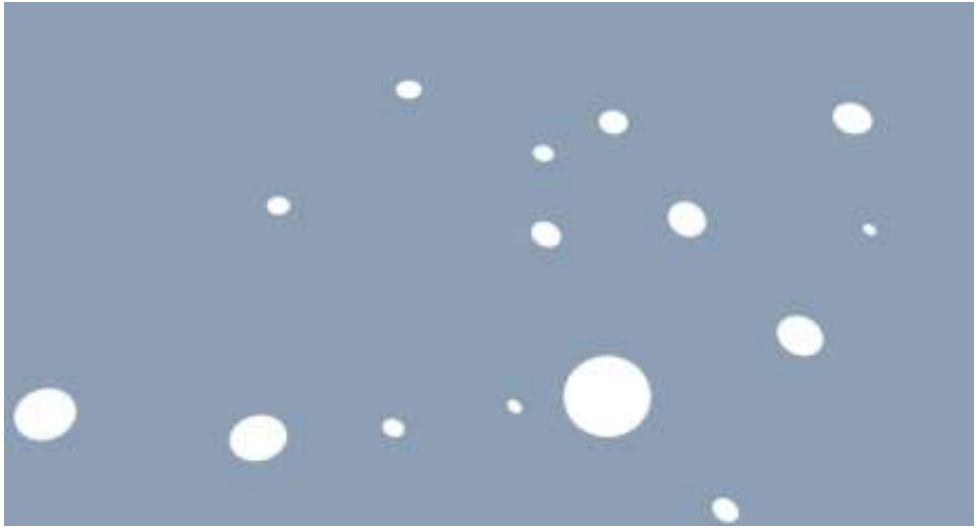

Fig.2: Oblique view of the surface of the liquid with large discs formed by small clusters of bubbles. The capillary forces, strong at short distances, are responsible for the cohesion of each circular aggregate, just as celestial objects such as spherical stars maintain themselves through their internal gravitation. At greater distances the capillary forces between the clusters are smaller, just as the star-to-star gravitational forces keep them inside galaxies. Each disc is surrounded by a meniscus (weak 3D deformation of the 2D surface, not seen here) the overlap of which is at the origin of capillary interactions. This figure, issued from a fractal-like distribution of the type of Fig. 1, is perfectly reproducible (only the size distribution of the clusters and their diameters depend on the initial conditions). Clearly, the discs of Fig.2 interact with each other and their interactions depend on their distance according to the capillary expression (given above) resulting from the overlap of their 3D meniscuses.If an initial velocity was given to one of them, it would "gravitate" ("capillarate") around others, as for stars, planets and other celestial objects.

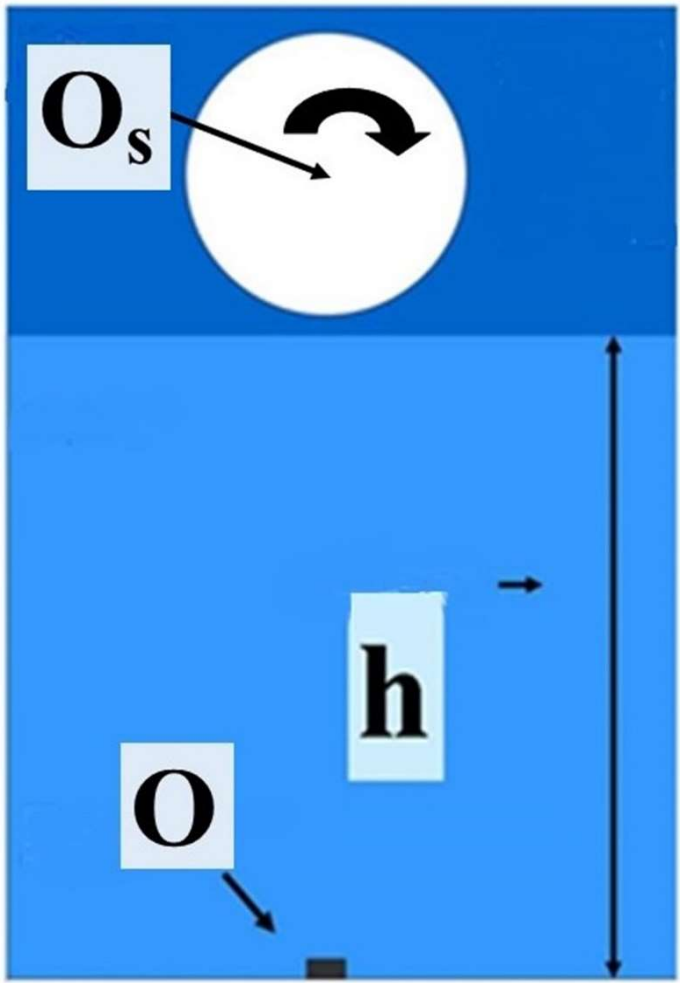

Fig 3 : An opening is created at point O at the bottom of the tank (small black square) with a depth of liquid h. The liquid is represented in blue, in its thickness, while its surface is represented in the same plane in dark blue. The white disk represents an initial cluster of small bubbles that congregate on the surface around point Os (vertical of the hole in O)

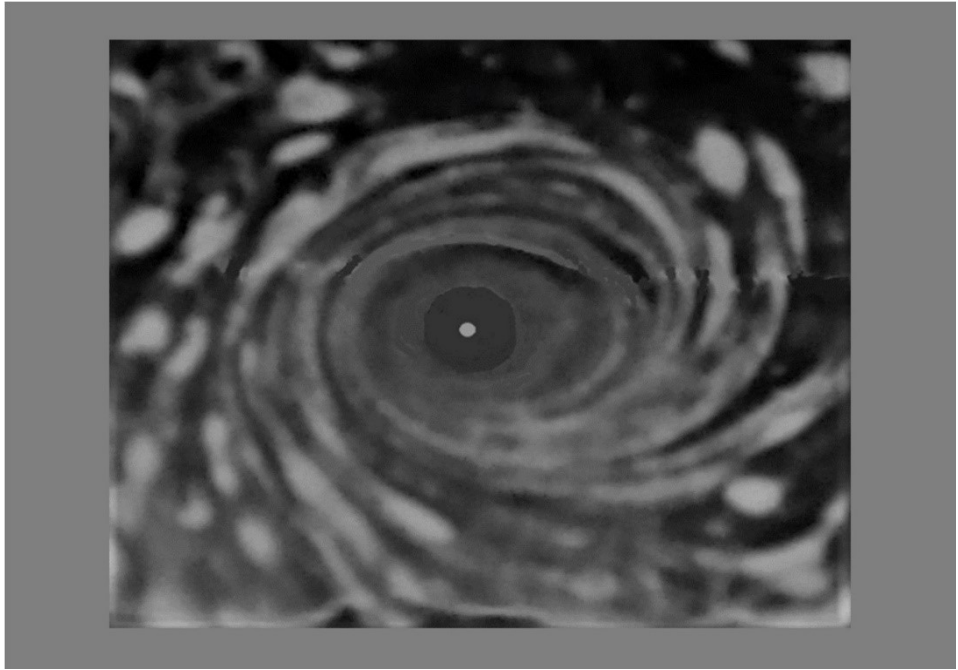

Fig. 4: A "bubble black hole" obtained with a sufficiently large density of bubbles and clusters of bubbles when their rotation velocity is sufficiently large. Explained in the next section, this happens when the distance $h$ between the hole, at the bottom of the tank, and the surface is small enough (top-right image of Fig. 8c). The white center and its gray halo represent a U2D "black hole". The picture is taken from above the liquid surface up hole. At a sufficiently long distance from the center (near the edges of the figure) a 2D "spiral galaxy" rotates around this "black hole".

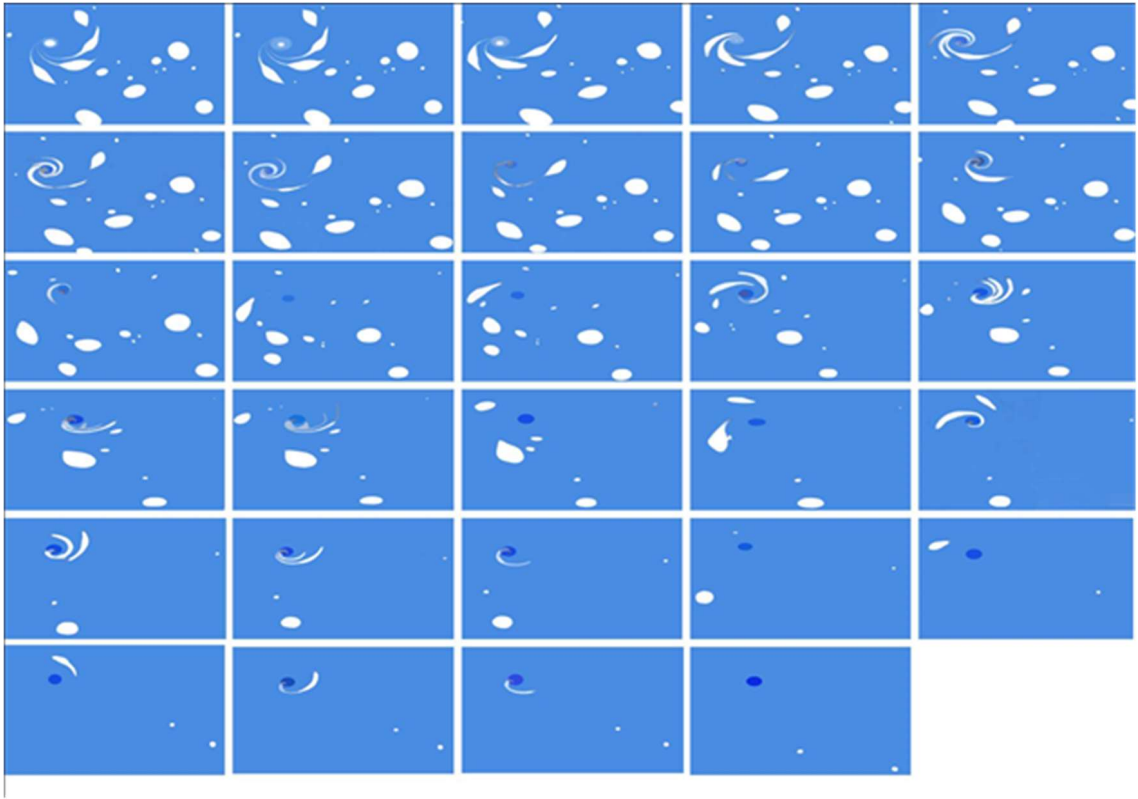

Fig. 5: View of bubble clusters rotating around the Os point on the surface of the liquid when the high *h* of the liquid is decreasing. This point is at the vertical direction of point O at the bottom of the water tank (not seen) where the hole is located. The whiter the area is, the greater the density of bubbles. The bubble clusters closest to Os begin by swirling around this point —clockwise here— while lengthening and then being "peeled", contributing to the formation of a "black hole halo" and later to the central disc of the "black hole" (see below). The rotation of the bubbles and clusters of bubbles faster and faster when the high *h* of water of the tank decreases. The way a black hole absorbs matter in our U3D is very similar to that shown for several published pictures or simulations

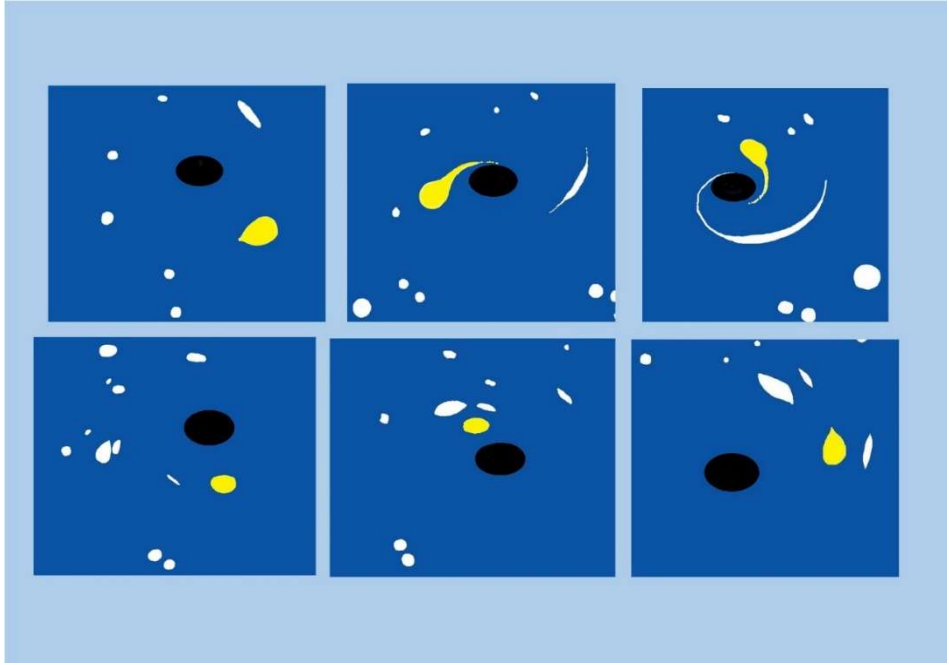

Fig. 6: During its gravitation ("cappileration") around point Os, the yellow bubble starts peel and after it loses enough matter (its tail) it is able to escape from the attraction of the "black hole" (entered at point Os). This is a 2D realization of the Penrose process allowing the extraction of energy from the black hole. In the last picture the yellow star is larger because it merges with another, smaller, star.

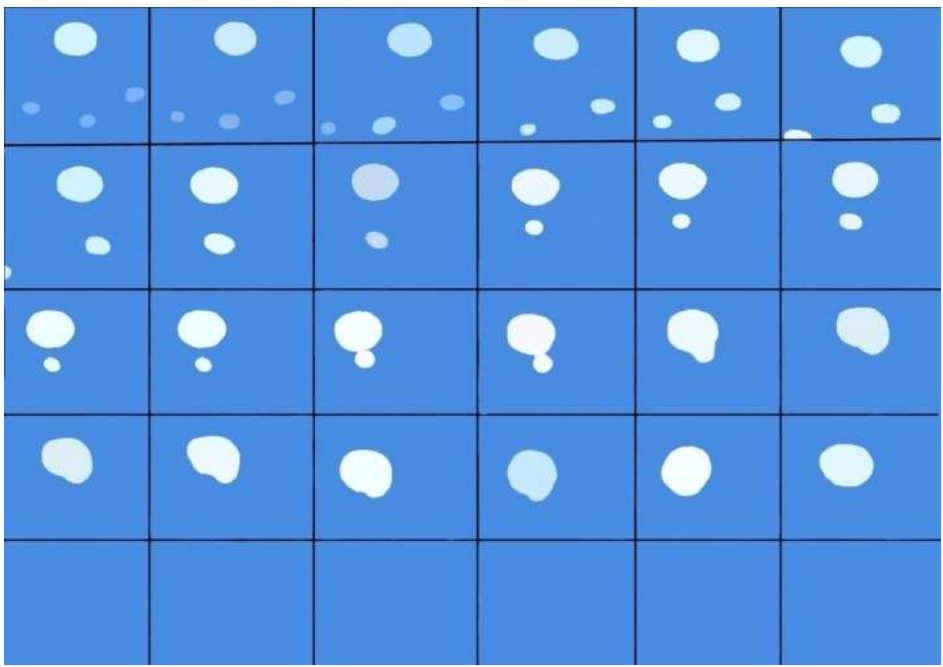

Fig. 7: Analogue in U2D of the progressive transformation of a double star into a single star in U3D, in rotation far away from the "black hole" (the largest star attracts a smaller star, on its way to the point Os, not seen).

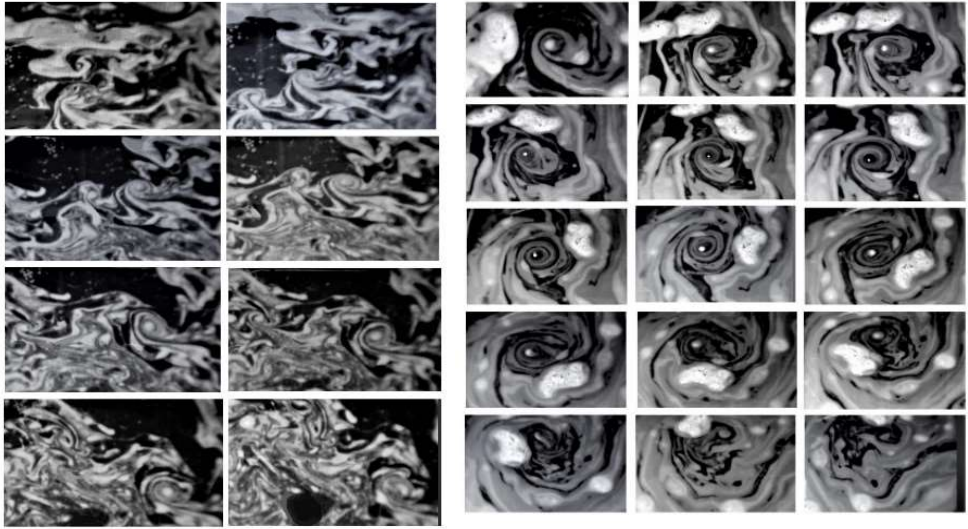

Fig. 8: Spontaneous creation of an "irregularly bubble galaxy" appearing progressively in the absence of a hole in the tank after mixing the water/bubble mixture and its progressive transformation after collision with a random ensemble of bubbles and clusters (left). The white center reflects a simple accumulation of bubbles analogous to a large star. The figure on the right corresponds to the case where whirls result from a hole at the bottom of the tank, with $h>h_i$, i.e. when the surface whirls are not strongly connected with the hole aspiration. Nevertheless, this hole produces whirls analogous to those of a real "black hole" (see next section) with its galaxy, but here the galaxy is moving. As in the case without a hole (left) this "galaxy" disappears after the collision with moving masses of bubbles. However, the hole being there, the collective rotation of the "galaxy" reappears after such a moving mass of bubbles has drifted away, mimicking the disappearance/appearance of a weak ($h>h_i$) black hole.

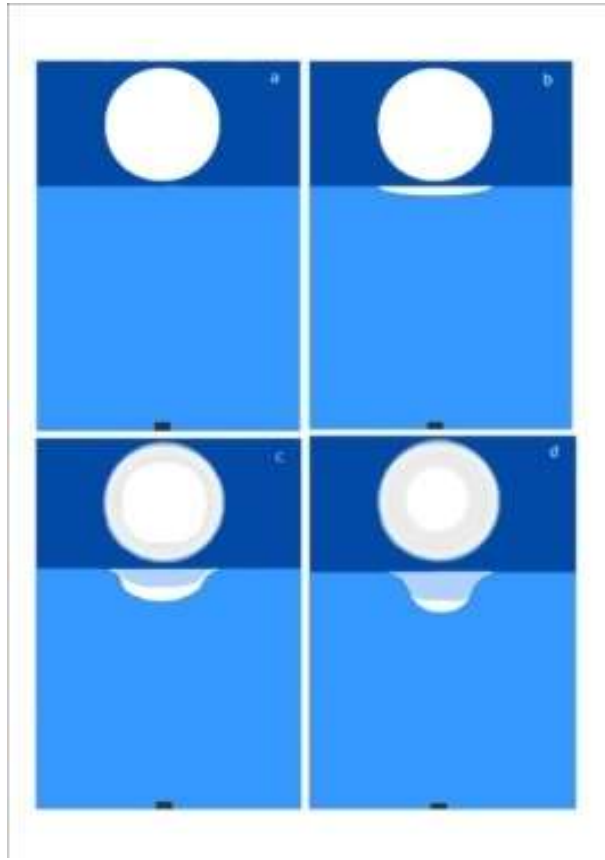

Fig. 9a: From left to right, top and then bottom. An opening is created at point O at the bottom of the tank (small black square) with a depth of liquid $h=h_i$. The liquid is represented in blue, in its thickness, while its surface is represented in the same plane in dark blue. The white disk represents an initial cluster of small bubbles that congregate on the surface around point Os (vertical of the hole in O) and are rapidly rotating. This cluster is very bright because of its high density of bubbles, which reflect light. As $h$ decreases halos are formed, they becomes darker and darker, corresponding to a lower density of bubbles. Indeed, the bubbles are transferred from the halo to the central disc. The observations show that the speed of rotation of this halo is smaller than that of the central disc, but both increase when $h$ decreases. Such central discs with their halo represent our 2D analog of a black hole of our universe, with its accretion disc. The optical observations, pictures and movies, were taken from the surface only. The representation of the location of the bubbles in the thickness of the liquid, deduced from these surface observations, is only a schematic, but enough to give a qualitative picture of what is going on (the lateral dimensions of the tank length were : 40 cm x 40 cm and its height 18 cm. The size of the hole was 0.7 cm).

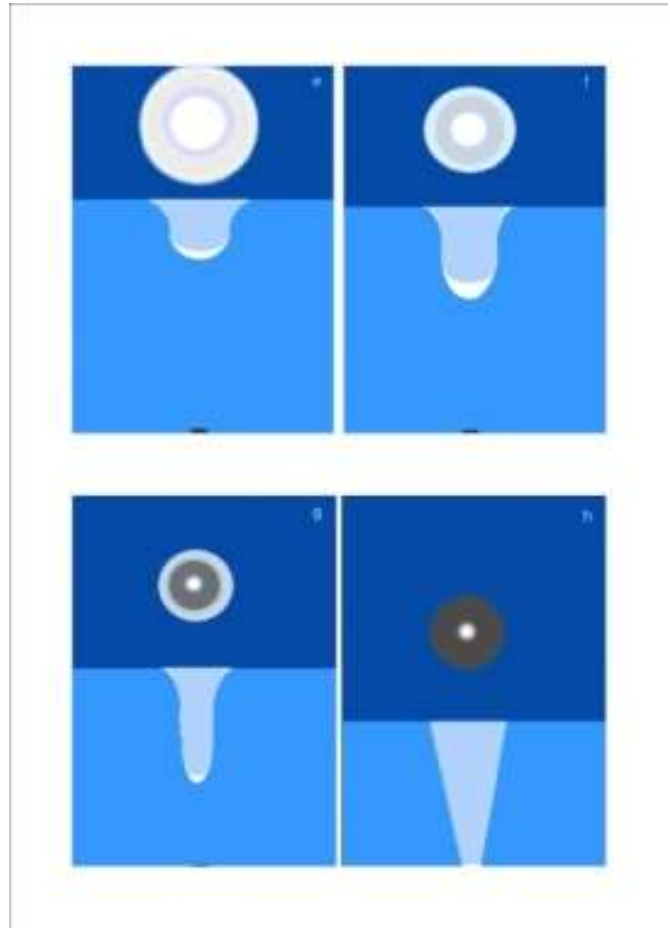

Fig. 9b: From left to right, top and then bottom. This is a follow up of the previous figure in which the liquid height $h$ continues to decrease. The regions in lightest blue are inside the tourbillon, with less liquid and splashes (not seen, but easy to imagine). Here one can see more clearly why the halo around the central disk becomes darker and darker (until becoming black) and larger and larger: it corresponds to what is seen from the surface of the walls of the "bowl" where the bubbles are located. When $h$ decreases, the density of bubbles decreases because the bowls walls become increasingly deeper and more and more vertical. The lightest central part is white because this is where most bubbles accumulate, at the bottom of the bowl, below the surface. Note that, for more visibility, this figure exaggerates the actual depth of the "bowl" that we could not measure precisely.

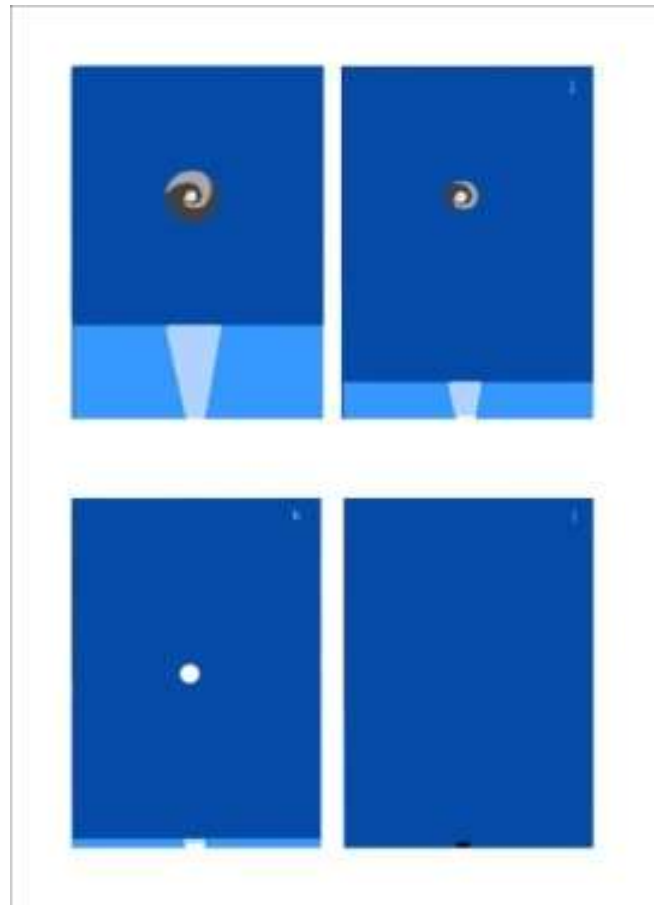

Fig. 9c: From left to right, top and then bottom. This is a follow up of the previous figure in which the liquid height *h*, continues to decrease. The halo is now black with a clearer gray curl meaning that the bubbles, previously sitting on the bowl edges, are completely evacuated by a tourbillon in the hole. The latter, white with the last bubbles, disappears after all the bubbles have been evacuated.                                                                                                                                                                                                                                                                                                In the static representation of Fig. 8, the central part, the "bowl" is always more or less empty of fluid (light blue). This representation should be completed by a short comment regarding dynamical aspects: when *h* decreases the fluid falls in the bowl more and more rapidly, in an increasingly non-uniform way so that the bowl may be, from time to time, empty (nothing to see) or with the fluid carrying bubbles (visible). This suggests that if a black hole carrying fluid and matter can be observed, it is not excluded to lose it from time to time.